\documentclass[aps,prl,twocolumn,floats,final,superscriptaddress,citeautoscript,longbibliography]{revtex4}
\usepackage{amssymb}
\usepackage{graphicx}
\usepackage{amsmath}
\usepackage{braket}
\usepackage[colorlinks=true,linkcolor=blue,citecolor=blue,urlcolor=blue]{hyperref}

\begin{document}

\title{Quantum Many-Body Scars and Quantum Criticality}
\author{Zhiyuan Yao}
\thanks{They contribute equally to this work. }
\affiliation{Institute for Advanced Study, Tsinghua University, Beijing 100084, China}
\author{Lei Pan}
\thanks{They contribute equally to this work. }
\affiliation{Institute for Advanced Study, Tsinghua University, Beijing 100084, China}
\author{Shang Liu}
\email{sliu.phys@gmail.edu}
\affiliation{Department of Physics, Harvard University, Cambridge, MA, 02138, USA}
\author{Hui Zhai}
\email{hzhai@tsinghua.edu.cn}
\affiliation{Institute for Advanced Study, Tsinghua University, Beijing 100084, China}
\date{\today}

\begin{abstract}

In this letter, we study the PXP Hamiltonian with an external magnetic field that exhibits both quantum scar states and quantum criticality. It is known that this model hosts a series of quantum many-body scar states violating quantum thermalization at zero magnetic field, and it also exhibits an Ising quantum phase transition driven by finite magnetic field. Although the former involves the properties of generic excited states and the latter concerns the low-energy physics, we discover two surprising connections between them, inspired by the observation that both states possess log-volume law entanglement entropies. First, we show that the quantum many-body scar states can be tracked to a set of quantum critical states, whose nature can be understood as pair-wisely occupied Fermi sea states. Second, we show that the partial violation of quantum thermalization diminishes in the quantum critical regime. We envision that these connections can be extended to general situations and readily verified in existing cold atom experimental platforms.

\end{abstract}

\maketitle

Thermalization lies at the center of statistical mechanics, and the eigenstate thermalization hypothesis (ETH) lays the foundation of quantum statistical mechanics \cite{Deutsch:1991ju,Srednicki:1994dl,Rigol:2008bf,DAlessio:2016gr}.
The ETH states that a highly excited eigenstate of a generic interacting quantum many-body system locally behaves as a thermal ensemble. A direct consequence of the ETH is that the entanglement entropy of a subsystem equals its thermal entropy and, therefore, displays the volume law behavior. So far, most quantum many-body systems are known to obey ETH, with few classes of exceptions. For example, the ETH can be violated by fine-tuning the system parameters to exactly solvable points or by adding strong disorder to make the system many-body localized \cite{Gornyi:2005fv,Basko:2006hh,Abanin:2019dl}.
\begin{figure}[t]
    \centering
    \includegraphics[width=0.4\textwidth]{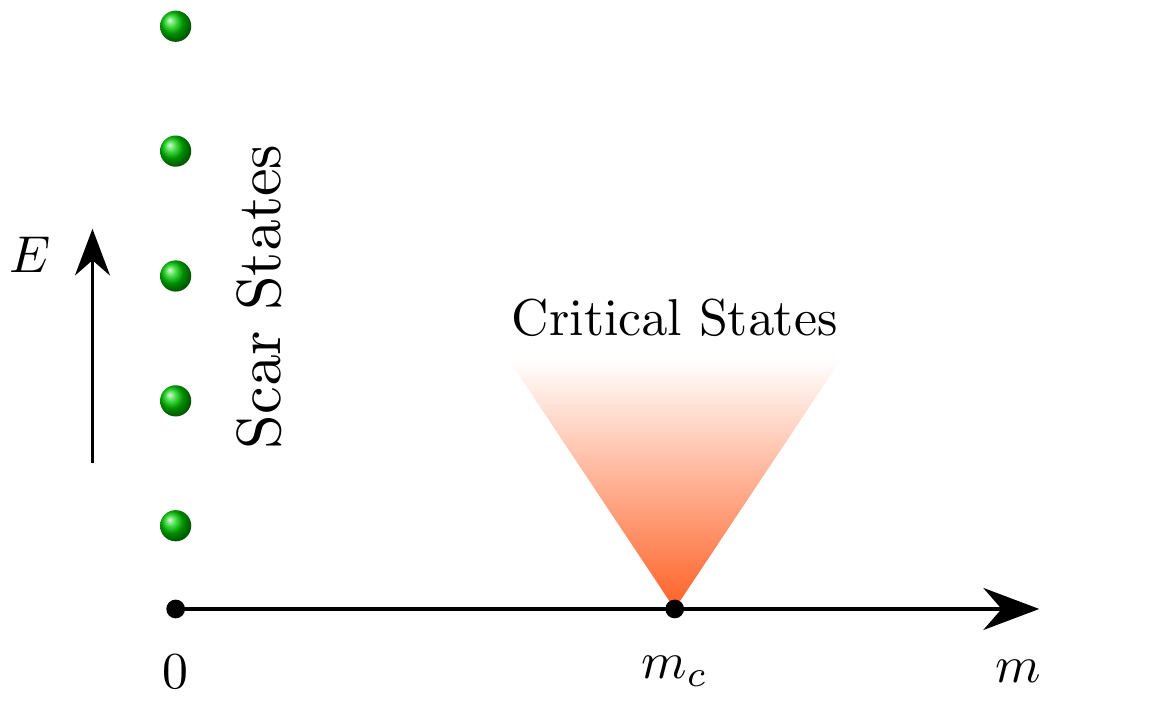}
    \caption{Schematic of two kinds of log-volume law states in the PXP model with an external magnetic field. One consists of quantum many-body scar states (green balls) first discovered at $m=0$, and the other one involves the low-energy critical states at the Ising quantum critical point $m_\text{c}$. Here $E$ denotes the energy and $m$ is the strength of the external field. This paper focuses on the relation between these two sets of quantum states.}
     \label{schematic}
\end{figure}

Recently a new mechanism of ETH violation has been discovered first in a Rydberg atom quantum simulator, where certain initial states are found to retain initial local information after sufficiently long time evolution \cite{Bernien:2017bp}. Later it is pointed out theoretically that such systems can be described by the so-called PXP model \cite{Turner:2018iz,Turner:2018in}. The Hamiltonian of this one-dimensional model reads
\begin{equation} \label{PXP}
	\hat{H}=\sum\limits_{i}\hat{P}_{i-1}\hat{\sigma}^x_i\hat{P}_{i+1}.
\end{equation}
In the Rydberg atom system, $\hat{\sigma}^x_i$ couples an atom at site-$i$ between the ground state $|g\rangle$ and the Rydberg excited state $|e\rangle$. Here $\hat{P}_i=(1-\hat{\sigma}^z_i)/2$ with $\hat{\sigma}_{i}^{z} = | e \rangle \langle e | - |g \rangle \langle g | $ is a projection operator onto the ground state. The PXP model is an effective model for the Rydberg blockade when the blockade radius is around one lattice spacing. This means that excitation to the Rydberg state at a given site is allowed only when atoms in its two neighboring sites are both in their ground states. Hence, in addition to the Hamiltonian Eq. \ref{PXP}, there exists an extra constraint that forbids two neighboring sites being both occupied by the Rydberg excited state, \emph{i.e.}, for any physically allowed state $|\Psi\rangle$ and for every site-$i$, it requires   
\begin{equation}
(1-\hat{P}_i)(1-\hat{P}_{i+1})|\Psi\rangle=0. \label{constraint}
\end{equation}

There are two unique aspects of quantum thermalization in the PXP model. First, violation of the ETH in this system neither requires fine-tuning parameters nor adding disorder. Second, ETH is only partially violated in this system. That is to say, only a subset of total eigenstates violate ETH, and the rest of eigenstates still obey ETH. These non-thermal states are called the quantum many-body scar states, which has attracted considerable attentions recently \cite{Lukin,Lukin2,SXu2019,Lukin2019PRL,Motrunich2019,Papic2020PRB,PXPgauge,Papic2021PRX,Hsieh2020,LukinScience2021,Papic2021Review}. Later, quantum many-body scar states have also been discovered in several different models \cite{Moudgalya:2018gt,Moudgalya:2018gj,Moudgalya:2020ip,Moudgalya:2020ez,Mark:2020dg,Mark:2020fp,Schecter:2019db,Iadecola2020,Katsura2020,WWHo2020,Lee2020,SGA1,SGA2,SGA3}, such as the Affleck--Kennedy--Lieb--Tasaki (AKLT) model, the spin-1 XY model, and the generalized Fermi--Hubbard model.
\begin{figure*}[htbp]
	\centering
	\includegraphics[width=0.9\textwidth]{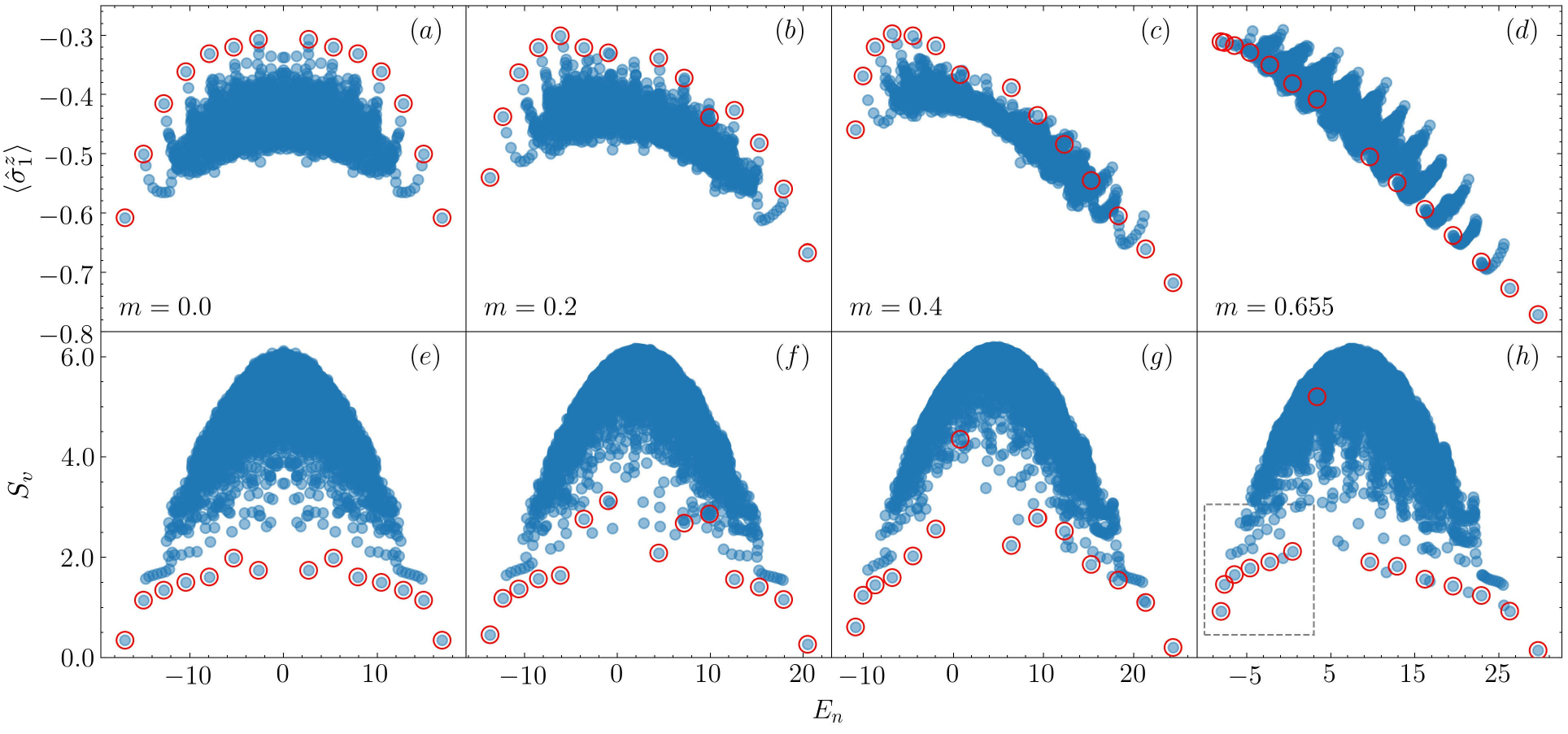}
	\caption{Snapshots of the process of tracking the quantum many-body scar states (highlighted by red circles) from $m=0$ to $m=m_\text{c}$. Here we present results for system size $L=28$ in the system sector $(k,I)=(0,+)$ at four typical values of $m$. For a given $m$, the top panel shows expectation values of local observables $\langle \hat{\sigma}^z_1\rangle$, and the corresponding bottom panel displays the bipartite von Neumann entanglement entropies (see text for definition) of all eigenstates.}
	\label{fig:tracking}
\end{figure*}

The entanglement entropies of these non-thermal scar states display the log-volume law behavior \cite{Turner:2018iz,Turner:2018in,Moudgalya:2018gj,Motrunich2019,Vafek:2017bv,Schecter:2019db,SGA2}. This is reminiscent of quantum critical states at a $(1+1)$D Ising critical point, whose entanglement entropies also display the same behavior \cite{Vidal:2003cn,Calabrese:2004hl}. To investigate possible connections between these two kinds of states, we need a model that exhibits both scar states and critical states. To this end, we consider adding an external magnetic field into the PXP model whose Hamiltonian reads 
\begin{equation} \label{PXP-m}
	\hat{H}=\sum\limits_{i}(\hat{P}_{i-1}\hat{\sigma}^x_i\hat{P}_{i+1}-m\hat{\sigma}^z_i) \, ,
\end{equation}
and we focus on the $m>0$ side. The external magnetic field can drive a quantum phase transition at $m=m_\text{c}\approx 0.655$ \cite{Note0,Sachdev:2002jl,Fendley:2004bw,Rico:2014ik}. When $m>m_\text{c}$, the external magnetic field tends to polarize all atoms into the $|e\rangle$ states. However, due to the Hilbert space constraint, at most a half of the atoms can be in the $|e\rangle$ states.
Moreover, in the limit of $m\rightarrow\infty$, the ground states are two-fold degenerate, which are $|gegege\dots\rangle$ and $|egegeg\dots\rangle$, and each of them breaks the $Z_2$ symmetry. Thus, this quantum phase transition belongs to the Ising universality class.

Hence, both scar states and critical states exist in the phase diagram of this model, as shown in Fig. \ref{schematic}. On the one hand, it has been well studied that the PXP model at $m=0$ hosts a set of approximately equally spaced scar states, ranging from the lowest to the highest energy of the model. On the other hand, around $m=m_\text{c}$, the presence of the Ising quantum phase transition ensures a set of low-energy quantum critical states governed by the Ising conformal field theory (CFT). Both states display the log-volume law entanglement entropies but due to different mechanisms.  In this letter, we address the question that whether there exists a connection between these two sets of seemingly different states, and more generally, whether there exists a connection between quantum thermalization and quantum criticality.

\textit{Tracking Scars to Criticality.} Considering a set of eigenstates $\{|n\rangle\}$ of the model Eq.~\eqref{PXP-m} with a given $m$, we then change $m$ to $m+\delta m$, and the set of eigenstates become $\{|n^\prime\rangle\}$. Now for a given eigenstate $|n_0\rangle$ in $\{|n\rangle\}$, we can find out a unique state $|n_0^\prime\rangle$ in $\{|n^\prime\rangle\}$, which maximizes $|\langle n^\prime_0|n_0\rangle|$. Then $|n^\prime_0\rangle$ is identified as the corresponding state of $|n_0\rangle$. In this way, we can keep tracking an eigenstate as $m$ varies. Here we focus on the situation that $|n_0\rangle$ is a scar state and we track the scar states from $m=0$ to the critical point $m=m_\text{c}$.

We numerically diagonalize the PXP model of system size $L=28$ under the periodic boundary condition. Besides the translational symmetry, the system also has a bond inversion symmetry, and the good quantum numbers associated with them are momentum $k$ and parity $I$.
Quantum many-body scar states exist only in $(k,I)=(0,+)$ and $(k,I)=(\pi,-)$ sectors where $+$ and $-$ denotes inversion even and odd respectively. Our results for the $(k,I)=(0,+)$ sector are shown in Fig.~\ref{fig:tracking}, and the results for the $(k,I)=(\pi,-)$ sector are similar,
For this state tracking, we have chosen $\delta m=0.1$ starting from $m=0$, and for most steps, the maximal overlap is around $0.9$.

As one can see from Fig.~\ref{fig:tracking}, there are two prominent features in this tracking process. First, as regards the energy, around half of the scar states move towards lower energies when $m$ approaches $m_\text{c}$.
Second, as regards the entropy, we employ the von Neumann entanglement entropy $S_v=-\text{Tr}_{A} \rho_A \log \rho_A$, where $\rho_{A}$ is the reduced density matrix of the subsystem $A$ after tracing out the rest of the system. In Fig.~\ref{fig:tracking}(e-h), we plot $S_v$ when $A$ is taken as half of the entire system, which is also called the bipartite entanglement entropy, for all eigenstates in the $(k, I) = (0, +)$ sector. It is clear that in the process of tracking to the critical point, the tracked scar states always retain entanglement entropies much lower than the surrounding thermal states. 

\begin{figure}[t]
    \centering
    \includegraphics[width=0.43\textwidth]{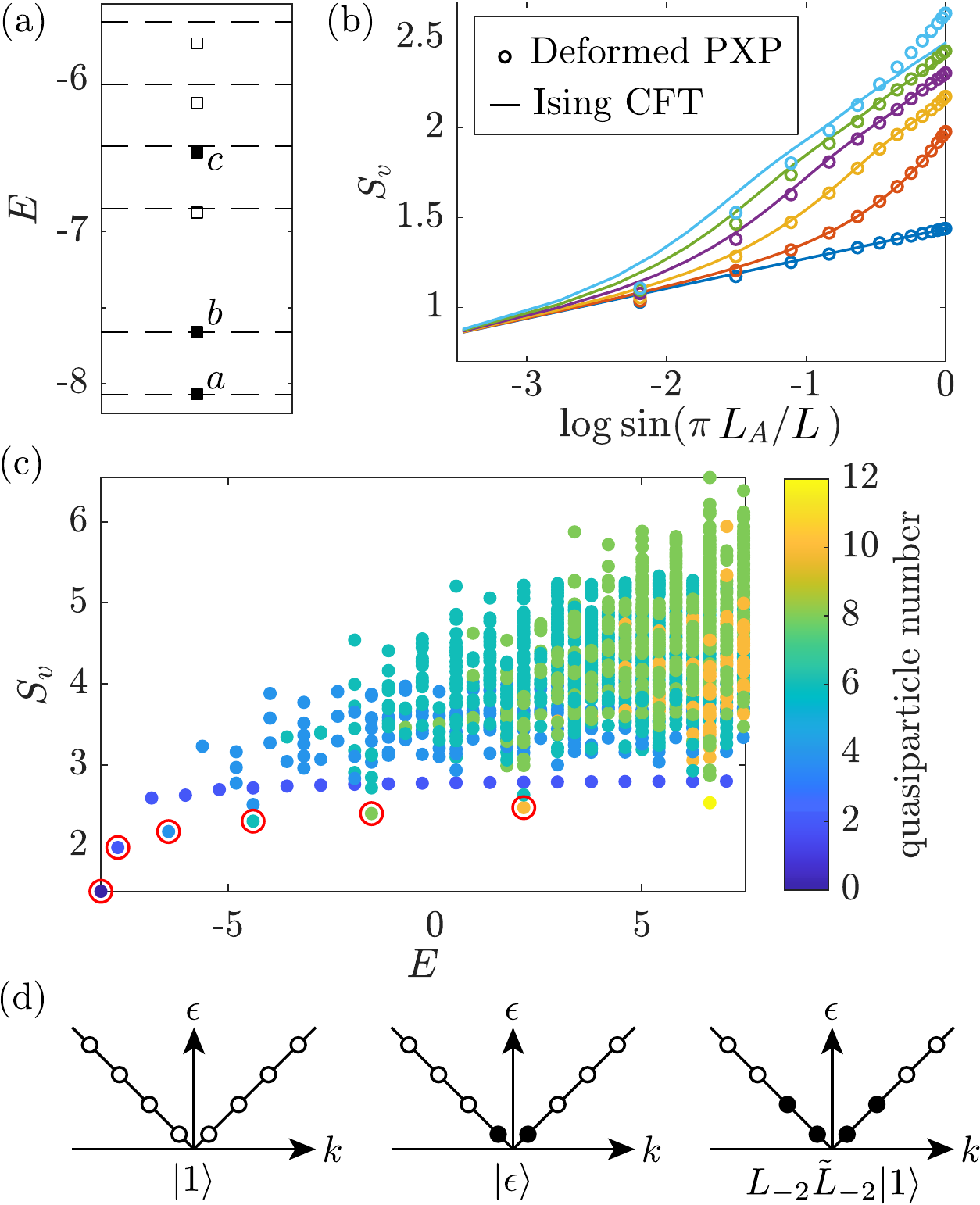}
    \caption{(a) Comparison between the low-energy spectrum of the PXP model with an external magnetic field at the criticality (boxes) and the prediction of the Ising CFT (dashed lines). Filled boxes represents the lowest three tracked scar states. (b) $S_v$ as a function of the size $L_A$ of the subsystem $A$, scaled by the total number of sites $L$. The discrete data points correspond to the tracked scar states in the dashed box of Fig.\,\ref{fig:tracking}(h). The solid lines correspond to the six low-energy and low-entropy states marked by red circles in (c), and are obtained from the free fermion theory of the transverse field Ising model. (c) The bipartite entanglement entropy for the transverse field Ising model. The color bar indicates the number of fermion quasiparticles of each state. (d) Illustration of the lowest three states marked by red circles in (c) (also labeled as $|a\rangle$, $|b\rangle$, and $|c\rangle$ in (a)), which are pair-wisely occupied Fermi sea states of free fermions.}
     \label{criticality}
\end{figure}

\textit{Low-Entropy States at Criticality.} In the dashed box in Fig.~\ref{fig:tracking}(h), we highlight a set of low-energy states at the critical point $m=m_\text{c}$, in which the states marked by red circles are the tracked scar states.
We shall show the nature of these scar states can be understood from the perspective of quantum criticality utilizing the Ising conformal field theory (CFT) \cite{Francesco:2012fo,Ginsparg:1988vv}. Below we will compare both the energies and the entanglement entropies of these tracked states at $m=m_\text{c}$ with the prediction of the Ising CFT. To make a proper comparison, we should caution that the symmetry breaking phase in the PXP model is an anti-ferromagnetic phase. Therefore, translation of one lattice spacing in the PXP model is related to the Ising $Z_2$ symmetry in the Ising CFT, and only translation of two lattice spacings in the PXP model corresponds to the translational symmetry in the Ising CFT. Similarly, the bond-center inversion in the PXP model is equivalent to the inversion in the Ising CFT followed by the action of the Ising $Z_2$ generator.

\begin{figure}[t]
    \centering
    \includegraphics[width=0.45\textwidth]{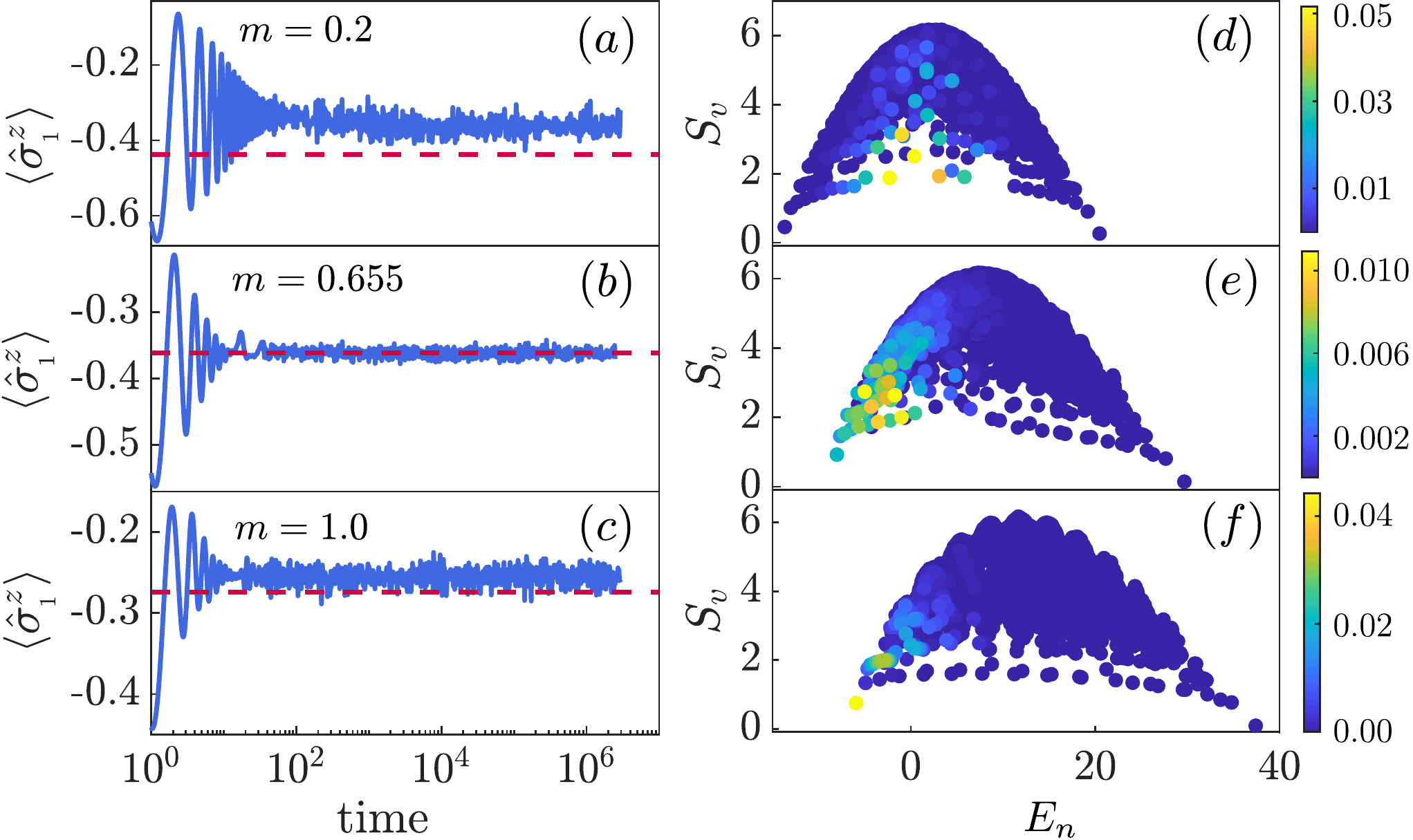}
    \caption{(a-c): Time evolution of the local observable $\langle \hat{\sigma}^z_{i} \rangle$ starting from the $\left|\mathbb{Z}_{2}\right\rangle$ initial state for different $m$. The dashed lines indicate the corresponding thermal averaged values. (d-f): The bipartite entanglement entropies of all eigenstates, with color bar representing the overlaps of the eigenstates with the $\left|\mathbb{Z}_{2}\right\rangle$ initial state. Here $m=0$ for (a) and (d), $m=m_\text{c}\approx 0.655$ for (b) and (e), and $m=1.0$ for (c) and (f). }
     \label{thermalization}
\end{figure}

Firstly, in Fig.~\ref{criticality}(a) we compare the energies of these tracked scar states with the spectrum of the Ising CFT \cite{Note1}, with the latter properly shifted and rescaled. We find perfect agreement at least for a few lowest energy states, including three tracked scar states denoted by $|a\rangle$, $|b\rangle$,  and $|c\rangle$ shown in Fig.~\ref{criticality}(a). In terms of the standard notation of the Ising CFT, they are identified as $|1\rangle$, $|\epsilon\rangle$, and $L_{-2}\tilde{L}_{-2}|1\rangle$, respectively \cite{Note2}. However, this comparison can hardly be extended to higher energy because the Ising CFT describes continuum models, and the finite size effect of the lattice model calculation will be more severe as energy increases.

Secondly, we consider the transverse field Ising model sitting at the Ising criticality. The transverse field Ising model and the PXP model share the same low-energy physics at criticality due to the universality of critical behaviors. The transverse field Ising model has the advantage that it can be written explicitly in terms of free fermions so that many physical quantities, such as the entanglement entropy, of the eigenstates can be analytically or semi-analytically computed \cite{Peschel:2003gz}. 
Here we properly choose the overall energy scale of the transverse field Ising model such that its spectrum matches that of the PXP model at the criticality. Then, we compute the bipartite entanglement entropy for the transverse field Ising model at the criticality with $L=10^3$ sites, as shown in Fig.~\ref{criticality}(c), where an overall vertical shift has been applied for comparison with the PXP model. As one can see, the entropy profile is similar to that of the PXP model shown in the dashed box of Fig.~\ref{fig:tracking}(h). In both cases, there exist a series of low-entropy states (marked by red circles). In Fig.~\ref{criticality}(b), we compare the entanglement entropies of these labeled states in the PXP model with the corresponding ones in the transverse field Ising model, and we obtain reasonably good agreements between them, both of which exhibit the log-volume law behavior. Therefore, these two sets of states share the same nature. Since the latter can be expressed in terms of the Slater determinants of free fermions, we can now reveal the nature of these tracked scar states. As shown in Fig.~\ref{criticality}(d), they are just pair-wisely occupied Fermi sea states of free fermions.

\textit{Thermalization of the $\mathbb{Z}_{2}$ State.} We have shown that the scar states are stable until reaching the critical point. One physical manifestation of these scar states is the violation of thermalization of the $\mathbb{Z}_2$ state. Here the $\mathbb{Z}_{2}$ state denotes the antiferromagnetic state $|gegege\dots\rangle$. This violation is attributed to the large overlap of the $\mathbb{Z}_2$ state with scar eigen-states \cite{Turner:2018iz,Turner:2018in}. Here we study how this violation of thermalization changes as the parameter $m$ varies across the critical point.
\begin{figure}[t]
    \centering
    \includegraphics[width=0.45\textwidth]{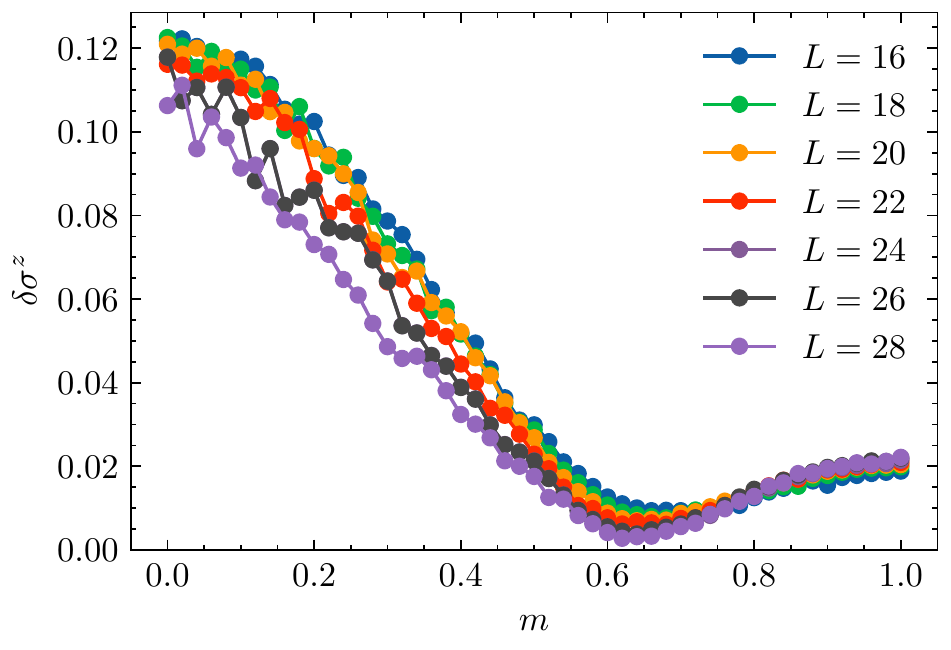}
    \caption{The difference $\delta \sigma^z$ between the long-time averaged local observable $\bar{\sigma}^z$ and the thermal equilibrium value $\sigma^{z}_{\text{th}}$ (see text for detailed definition) as a function of the magnetic field $m$, for several different system sizes $L$. }
     \label{thermal_critical}
\end{figure}

Starting from the $\mathbb{Z}_{2}$ initial state, we compute the time evolution of the local observable $\langle \Psi(t)|\hat{ \sigma}^z_i|\Psi(t)\rangle$ at site $i$ (below we take $i=1$ for concreteness), where $|\Psi(t)\rangle=e^{-i\hat{H}t}|\mathbb{Z}_{2}\rangle$.
As shown in Fig.~\ref{thermalization}(a-c), after a sufficiently long time, this local observable eventually fluctuates around a stationary value. The stationary value $\bar{\sigma}^z$ can be obtained by calculating the long-time average as
\begin{equation}
    \bar{\sigma}^z = \frac{1}{T}\int_{t_0}^{t_0+T}\langle \Psi(t)| \hat{\sigma}^z_i|\Psi(t)\rangle dt,
\end{equation}
where in practice we take both $t_0$ and $T$ as sufficiently long time scales. We compare $\bar{\sigma}^z$ with the thermal equilibrium value $\sigma^{z}_\text{th}$ defined as 
\begin{equation}
    \sigma^{z}_{\text{th}}=\text{Tr} \, \big( \rho_{\text{th}}\hat{\sigma}^z_1 \big) \, .
\end{equation}
Here the thermal density matrix $\rho_{\text{th}}$ is given by
\begin{equation}
\rho_{\text{th}} = \frac{1}{\mathcal{Z}}e^{-\beta\hat{H}-\lambda \hat{\Pi}}, \   \  \mathcal{Z}=\text{Tr} \, e^{-\beta\hat{H}-\lambda \hat{\Pi}},
\end{equation}
where $\beta$ is the inverse temperature and $\hat{\Pi}$ is the total momentum operator with $\lambda$ being the associated Lagrange multiplier. The values of $\beta$ and $\lambda$ are then determined by matching the energy and the momentum conservation conditions as $\langle \mathbb{Z}_{2}|\hat{H}|\mathbb{Z}_{2}\rangle=\text{Tr}\, ( \hat{H}\rho_{\text{th}} )$ and $\langle \mathbb{Z}_{2}|\hat{\Pi}|\mathbb{Z}_{2}\rangle=\text{Tr}\, (\hat{\Pi}\rho_{\text{th}})$. 
The computed values of $\sigma^{z}_{\text{th}}$ are shown by the dashed lines in Fig.~\ref{thermalization}(a-c). Then, we compute the quantity $\delta \sigma^z=\bar{\sigma}^z-\sigma^{z}_{\text{th}}$, and use this quantity to quantify the degree of thermalization violation.

The deviation $\delta \sigma^z$ as a function of $m$ is plotted in Fig.~\ref{thermal_critical}. Remarkably, when $m$ increases from zero, we find that $\delta \sigma^z$ first decreases before the critical point when $m<m_\text{c}$, then vanishes in the critical regime $m\approx m_\text{c}$, and finally increases again above the critical point when $m>m_\text{c}$.
This trend can also be seen in three typical situations in Fig.~\ref{thermalization}(a-c) for $m<m_\text{c}$, $m = 0.655 \approx m_\text{c}$, and $m>m_\text{c}$, respectively.

To understand this feature, we use color plot to illustrate the weight of the $\mathbb{Z}_{2}$ state on different eigenstates in Fig.~\ref{thermalization}(d-f), in which the bipartite entanglement entropies are used to distinguish the scar states from the thermal states. In Fig. \ref{thermalization}(d) and (e), we can see that the main weights of the $\mathbb{Z}_{2}$ state shift from the scar states to the thermal states when $m$ increases towards $m_\text{c}$. As a result, the deviation from thermalization becomes smaller.  When $m>m_\text{c}$, we can see from Fig.~\ref{thermalization}(f) that the weight of the $\mathbb{Z}_{2}$ state becomes more and more concentrated on the ground state, because the $\mathbb{Z}_{2}$ state is one of the symmetry breaking ground states in the large $m$ limit. Consequently, the deviation from thermalization increases again because the ground state is usually a non-thermal state. That is to say, the $\mathbb{Z}_{2}$ state does not thermalize at the $m<m_\text{c}$ side because of its large overlaps with the scar states, and also does not thermalize at the $m>m_\text{c}$ side because of its large overlaps with the symmetry breaking ground state. Hence, the regime where the $\mathbb{Z}_{2}$ state thermalizes coincides with the quantum critical regime.  

\textit{Summary.} In summary, we study the PXP model with an external Zeeman field, in which quantum many-body scar states have been discovered at $m=0$ and an Ising quantum critical point exists at $m=m_\text{c}\approx 0.655$. In this work, we find two connections between the quantum many-body scar states and the quantum criticality. First, when a set of scar states are tracked to the critical point, they become low-energy states and, meanwhile, retain low entanglement entropies violating the volume law. Moreover, the Ising critical theory offers a clear physical meaning of these tracked states at the critical point as pair-wisely occupied Fermi sea states.
Second, under sufficiently long time evolution, the $\mathbb{Z}_{2}$ state thermalizes only in the critical regime. We highlight that the connections between quantum thermalization and quantum criticality are highly nontrivial. This is because the quantum thermalization concerns the properties of generic excited states, and, on the contrary, the quantum criticality concerns the low-energy behavior. They are not expected to be correlated at first glance. For future studies, on the theory side, we envision that this connection discovered in the PXP model can be extended to other more general situations. On the experimental side, our prediction can be directly verified in recent Rydberg atom experiments described by the PXP Hamiltonian. It can also be verified in the recent cold atom simulation of a lattice gauge model, whose effective Hamiltonian is mathematically equivalent to the PXP Hamiltonian \cite{Surace:2020dy,Cheng:2021bh}.

\textit{Acknowledgment.} We thank Pengfei Zhang and Yanting Cheng for insightful discussions. This work is supported by Beijing Outstanding Young Scientist Program and NSFC Grant No. 11734010.

\end{document}